\documentclass[aps,prb,twocolumn,superscriptaddress,showpacs]{revtex4-1}

\usepackage{graphicx}
\usepackage{amssymb}
\usepackage{dcolumn}
\usepackage{textcomp}
\usepackage{amsmath}

\renewcommand{\Re}{{\rm Re}}

\begin{document}

\title[Dependence of MQT rate on JJ area]{Dependence of the Macroscopic
Quantum Tunneling Rate on Josephson Junction Area}

\author{Christoph~Kaiser}
\affiliation{Institut f{\"u}r Mikro- und Nanoelektronische Systeme,
Karlsruher Institut f\"ur Technologie, Hertzstra{\ss}e 16, D-76187
Karlsruhe, Germany}

\author{Roland~Sch\"afer}
\email[]{Roland.Schaefer@kit.edu}
\affiliation{Institut f\"ur Festk\"orperphysik, Karlsruher Institut f\"ur
Technologie, Hermann-von-Helmholtz-Platz 1, D-76344
Eggenstein-Leopoldshafen, Germany} \affiliation{Center for Functional
Nanostructures, Karlsruher Institut f\"ur Technologie,
Wolfgang-Gaede-Stra{\ss}e 1a, D-76128 Karlsruhe, Germany}

\author{Michael~Siegel}
\affiliation{Institut f{\"u}r Mikro- und Nanoelektronische Systeme,
Karlsruher Institut f\"ur Technologie, Hertzstra{\ss}e 16, D-76187
Karlsruhe, Germany}
\affiliation{Center for Functional Nanostructures, Karlsruher Institut
f\"ur Technologie, Wolfgang-Gaede-Stra{\ss}e 1a, D-76128 Karlsruhe,
Germany}


\date{\today}

\begin{abstract}
We have carried out systematic Macroscopic Quantum Tunneling (MQT)
experiments on Nb/Al-AlO$_x$/Nb Josephson junctions (JJs) of different
areas. Employing on-chip lumped element inductors, we have decoupled the
JJs from their environmental line impedances at the frequencies relevant
for MQT.  This allowed us to study the crossover from the thermal to the
quantum regime in the low damping limit. A clear reduction of the crossover
temperature with increasing JJ size is observed and found to be in
excellent agreement with theory. All junctions were realized on the same
chip and were thoroughly characterized before the quantum measurements.
\end{abstract}

\pacs{74.50.+r, 85.25.Cp, 74.78.Na}

\maketitle

\section{Introduction}
Since the gauge-invariant phase over a Josephson junction (JJ) $\varphi$ is
a macroscopic variable, circuits containing JJs have been used as model
systems for the investigation of quantum dynamics on a macroscopic scale.
This research has recently led to the development of different types of
superconducting quantum bits \cite{Nakamura_first_charge_qubit,
Friedman_first_qubit, Martinis_first_current_biased_qubit,
Vion_hybrid_qubit, Chiorescu_first_flux_qubit}, which are promising
candidates for the implementation of quantum computers.  The starting point
of this field was the observation of Macroscopic Quantum Tunneling (MQT) in
Josephson junctions in the 1980s \cite{MQT_Voss_Webb,MQT_Martinis_1987}. In
such experiments, the macroscopic variable $\varphi$ is trapped in the
local minimum of a tilted washboard potential, before it tunnels through
the potential barrier and starts rolling down the sloped potential. Since
this running state is equivalent to the occurrence of a voltage drop over
the junction, such tunneling events can be experimentally detected.
MQT---often referred to as secondary quantum effect---is the manifestation
of the quantum mechanical behavior of a single macroscopic degree of
freedom in a complex quantum system. Furthermore, it is the main effect on
which all quantum devices operated in the phase regime (such as phase
qubits and flux qubits) are based. Consequently, the detailed understanding
of MQT is not only interesting by itself, but also important for current
research on superconducting qubits operated in the phase regime. In this
article, we report on a systematic experimental study of the dependence of
the macroscopic quantum tunneling rate on the Josephson junction area,
which to our knowledge has never been performed before. As
usual\cite{MQT_Voss_Webb, MQT_Martinis_1987, Wallraff2003}, we measure the
rate at which the escape of $\varphi$ out of the local minimum of the
washboard potential occurs as a function of temperature. At high
temperatures, the escape is driven by thermal fluctuation over the barrier
while it is dominated by tunneling at low temperatures. This leads to a
characteristic saturation of the temperature dependent tunneling rate below
a crossover temperature $T_\mathrm{cr}$, which is the hallmark of MQT. The
rates above and below crossover are affected by the dissipative coupling to
the environment of the JJ, which is commonly accounted for by a quality
factor $Q$ in theoretical descriptions. A major goal of the presented study
was to keep the influence of $Q$ on the rate constant while varying the
junction area, so that a change in the observed escape rates could be
clearly assigned to the changed JJ size.  For this purpose, we work in the
underdamped regime of large $Q$, which is only possible if the JJ is to
some extent decoupled from its low-impedance environment (i.e. the
transmission line leading to the JJ). We achieve this by employing on-chip
lumped element inductors.

This article is organized as follows: First, the physical model of MQT is
discussed and the theoretical expectations for varying junction size are
given. Second, the procedure and setup of measurement are described.
Afterwards, the investigated Josephson junctions are characterized
carefully, and finally, the results of the MQT measurements are presented
and discussed.

\section{Model and Macroscopic Quantum Tunneling}

\subsection{General Model}
The dynamics of a JJ is usually described by the RCSJ (resistively and
capacitively shunted junction) model \cite{RCSJ_Stewart,RCSJ_McCumber}. The
current flowing into the connecting leads comprises in addition to the
Josephson current $I_J=I_c\sin\varphi$ ($I_c$ denotes the critical current
of the junction) a displacement current due to a shunting capacitance $C$
and a dissipative component due to a frequency dependent shunting
resistance $R$. For a complete description, the electromagnetic environment
given by the measurement setup can be included in the model parameters. In
our case, $R$ will be influenced by the environmental impedance while $C$
can be regarded as solely determined by the plate capacitor geometry of the
JJ itself. In any case, the bias current $I$ is composed of
\begin{equation}\label{RCSJ}
I = I_c\sin \varphi + \frac{1}{R}\frac{\Phi_0}{2\pi}\dot\varphi +
C\frac{\Phi_0}{2\pi}\ddot\varphi\, ,
\end{equation}
where $\varphi$ is the gauge-invariant phase difference across the junction
and $\Phi_0=h/2e$ is the magnetic flux quantum.  The dynamics of $\varphi$
as expressed by (\ref{RCSJ}) is equally described by the well-studied
Langevin equation
\begin{equation}\label{Langevin}
M\ddot{\varphi}+\eta{}M\dot{\varphi}+\frac{\partial{}U}{\partial{}\varphi}=
\xi(t)\, ,
\end{equation} which describes a particle of mass $M=C(\Phi_0/2\pi)^2$ in a
tilted washboard potential 
\begin{equation}\label{potential}
U\left(\varphi\right) = E_J \left(1
- \cos\varphi - \gamma\varphi\right)\, ,
\end{equation}
exposed to damping $\eta=1/RC$ and under the influence of a fluctuating
force $\xi(t)$. The strength of $\xi(t)$ is linked to temperature and
damping by the fluctuation-dissipation theorem. Furthermore, $\gamma=I/I_c$
denotes the normalized bias current while $E_J=\Phi_0 I_c/2\pi$ is called
the Josephson coupling energy.  For $\gamma<1$, if thermal and quantum
fluctuations are ignored, the particle is trapped behind a potential
barrier
\begin{equation}\label{DeltaU}
\Delta U = 2E_J \left(\sqrt{1-\gamma^2} - \gamma\arccos\gamma \right)\, ,
\end{equation}
and the JJ stays in the zero-voltage state. In the potential well, the
phase oscillates with the bias current dependent plasma frequency
\begin{equation}\label{omegap}
\omega_p = \omega_{p0}\left(1-\gamma^2\right)^{1/4} = \sqrt{\frac{2\pi
I_c}{\Phi_0 C}} \left(1-\gamma^2\right)^{1/4}\, .
\end{equation}
To complete the list of important system parameters given in this section,
we introduce the quality factor
\begin{equation}\label{equ_Qdamp}
Q=\omega_p/\eta=\omega_pRC\, ,
\end{equation} which is conventionally used to quantify the damping in the
JJ.

At finite temperatures, the thermal energy $k_BT$ ($k_B$ being Boltzmann's
constant) described by $\xi(t)$ in (\ref{Langevin}) can lift the phase
particle over the potential barrier before the critical current $\gamma=1$
is reached, so that the particle will start rolling down the potential.
This is called premature switching and the observed maximal supercurrent
$I_\mathrm{sw}<I_c$ is called the switching current. When the phase
particle is rolling, the JJ is in the voltage state, since a voltage drop
according to $\dot{\varphi}=(2\pi/\Phi_0)V$ is observed. The thermal escape
from the potential well occurs with a rate \cite{Kramers1940,Haenggi1990} 
\begin{equation}\label{Gamma_th}
\Gamma_\mathrm{th}=a_t\frac{\omega_p}{2\pi}\exp\left(-\frac{\Delta
U}{k_BT}\right)\, ,
\end{equation}
where $a_t$ is a temperature and damping dependent prefactor, which will be
discussed in more detail in Sec.~\ref{sec_damping}. 

For $T\rightarrow 0$, where $\Gamma_\mathrm{th}\rightarrow 0$, premature
switching will still be present due to quantum tunneling through the
potential barrier. As the phase difference over the JJ is a macroscopic
variable, this phenomenon is often referred to as "Macroscopic Quantum
Tunneling" (MQT).  This means that by measuring the switching events of a
JJ for decreasing temperature, one will see a temperature dependent
behavior (dominated by the Arrhenius factor $\exp(-\Delta{}U/k_BT)$ in
(\ref{Gamma_th})) until a crossover to the quantum regime is observed. The
crossover temperature $T_\mathrm{cr}$ is approximately given by
\cite{Affleck_Tcr,MQT_Martinis_1987}
\begin{equation}\label{Tcr}
T_\mathrm{cr}=\frac{\hbar\omega_p}{2\pi k_B}=
\frac{\hbar\omega_{p0}}{2\pi{}k_B}\cdot(1-\gamma^2)^{1/4}\, ,
\end{equation}
where $\hbar$ is Planck's constant.  We can write the quantum tunneling
rate for temperatures well below crossover as
\cite{CaldeiraLeggett,MQT_Martinis_1987,Grabert87,Freidkin88}
\begin{equation}\label{Gamma_qu}
\Gamma_q=a_q\frac{\omega_p}{2\pi}\exp(-B)\, ,
\end{equation}
where $a_q=\sqrt{864\pi\Delta{}U/\hbar{}\omega_p}\exp(1.430/Q)$ and
$B=(36\Delta U/5\hbar\omega_p)(1+0.87/Q).$  In the limit of large $Q$, the
escape rate is expected to approach the temperature independent expression
(\ref{Gamma_qu}) quickly\cite{Grabert87,Freidkin86,*Freidkin87} once the
temperature falls below $T_\mathrm{cr}.$ The rates in (\ref{Gamma_th}) and
(\ref{Gamma_qu}) are functions of the normalized bias current $\gamma$ via
(\ref{DeltaU}) and (\ref{omegap}). The crossover to the quantum regime can
be nicely visualized by measuring the bias current dependence of the escape
rate $\Gamma(I)$ for a sequence of falling temperatures. The data are then
described over the whole temperature range by the thermal rate
(\ref{Gamma_th}) with the temperature as a fitting parameter.  In this way,
one obtains a virtual "escape temperature" $T_\mathrm{esc}$, which can be
compared to the actual bath temperature $T$. In the thermal regime, one
should obtain $T_\mathrm{esc}=T$ while in the quantum regime, one should
get $T_\mathrm{esc}=T_\mathrm{cr}=\mathrm{const}$.

\subsection{Influence of JJ Size on MQT}\label{sec_JJsize}
\begin{figure}[t] \centering
\includegraphics[width=\linewidth,angle=0,clip]{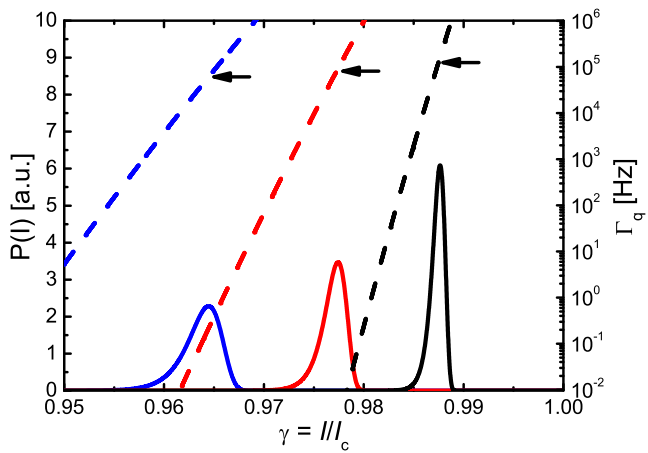}
\caption{\label{PI_rates}Theoretically calculated switching current
distributions $P(I)$ (solid curves) and quantum tunneling rates $\Gamma_q$
(dashed curves) for samples B1 (left) to B3 (right) having different
diameters $d$ (for parameters see Tab.~\ref{table_exp_Tcr}). The difference
in $\gamma_\mathrm{cr}$ (maximum position of $P(I)$), where quantum
tunneling leads to escape from the potential well, is significant. The
tunneling rates at these points (marked by arrows) are of the order of
$\approx100$~kHz for all samples.}
\end{figure}
The crucial element of a Nb-based Josephson junction as employed in this
work is the Nb/Al-AlO$_x$/Nb trilayer. For a JJ, $I_c=j_c\cdot A$ and
$C=c\cdot A$, where the critical current density $j_c$ and the specific
capacitance $c$ are constant for a given trilayer and $A$ is the area of
the junction. Hence, by reformulating (\ref{omegap}), we find that
$\omega_{p0}=\sqrt{2\pi j_c/\Phi_0 c}$, meaning that for JJs fabricated
with the same trilayer, the plasma frequency does not depend on their size.
So at first sight, the crossover temperature (\ref{Tcr}) should also be
independent of the JJ size. In reality, however, the problem is more
subtle, as one needs to take into account at which normalized bias current
$\gamma_\mathrm{cr}$ the quantum tunneling rate (\ref{Gamma_qu}) becomes
significant. Since the height of the potential barrier $\Delta U\propto
E_J\propto A$ is proportional to the JJ size,  a significant tunneling rate
should be reached at different $\gamma_\mathrm{cr}$ values for junctions of
different size. These points can be estimated by theoretically calculating
(\ref{Gamma_qu}) and converting it into a switching current histogram, as
it would be observed in a real experiment. The probability distributions of
switching currents $P(\gamma)$ can be obtained from the quantum rate by
equating \cite{Fulton_D}
\begin{equation}\label{PIfromGamma}
P(\gamma)=\Gamma_q\left(\frac{{\rm d}\gamma}{{\rm
d}t}\right)^{-1}\left(1-\int_0^I P(u){\rm d}u\right)\, ,
\end{equation}
where ${\rm d}\gamma/{\rm d}t=(1/I_c)\cdot({\rm d}I/{\rm d}t)$ is a
constant for the linear current ramp chosen in our experiment. For the
parameters of the junctions investigated in this work (see
Tab.~\ref{table_exp_Tcr}), the switching current distributions $P(\gamma)$
were determined with a quality factor of $Q=100$ and a current ramp rate of
100~Hz, according to our experiments (see below). They are shown in
Fig.~\ref{PI_rates}, where it can be seen that the $\gamma_\mathrm{cr}$
values (the positions of the maxima of the distributions) significantly and
systematically increase with the junction size. Evaluation of the
$\Gamma_q(\gamma_\mathrm{cr})$ values for the samples indicates that
quantum tunneling will be experimentally observable at a rate of around
$\Gamma_q\approx 10^5$~Hz.

Subsequently, the expected crossover temperature was calculated from
(\ref{Tcr}) with $\gamma=\gamma_\mathrm{cr}$.  The sample parameters as
well as the expected $\gamma_\mathrm{cr}$ and $T_\mathrm{cr}$ values are
given in Table~\ref{table_exp_Tcr}. It can be seen that due to the term in
parenthesis on the right hand side of (\ref{Tcr}), the crossover
temperature systematically decreases for increasing junction size. The
change in $T_\mathrm{cr}$ is large enough to be observed experimentally.
However, such a systematic study of the size-dependence of $T_\mathrm{cr}$
has never been carried out before.

\begin{table}[ht]
\caption{\label{table_exp_Tcr}Parameters of the investigated samples and
calculated $\gamma_\mathrm{cr}$ and $T_\mathrm{cr}$ values. The critical
current density accounts for $j_\mathrm{c}\approx 650$~A/cm$^2$ while the
specific capacitance of $c=55$~fF/\textmu{}m$^2$ was recently determined by
the measurement of Fiske steps on a trilayer prepared in our lab under
identical conditions. Parasitic capacitances due to idle regions next to
the JJs have been taken into account for all calculations in this article.
For the calculation of $T_\mathrm{cr}$, the actual measured critical
currents were used.}
\begin{ruledtabular} \begin{tabular}{cdcc}
Sample & \multicolumn{1}{c}{Diameter $d$ (\textmu{}m)} &
$\gamma_\mathrm{cr}$ &
$T_\mathrm{cr}$ (mK)  \\\hline
B1 & 1.9  & $0.965$ & $371$\\
B2 & 2.55 & $0.977$ & $323$\\
B3 & 3.6  & $0.988$ & $291$\\
B4 & 3.8  & $0.988$ & $277$\\ \end{tabular} \end{ruledtabular}
\end{table}

\subsection{Influence of Damping on MQT}\label{sec_damping}
The quality parameter $Q$ is frequently employed to describe the strength
of the hysteresis in the current-voltage characteristics of a JJ. In this
case, one often takes $Q=\omega_{p0}R_\mathrm{sg}C$ with $R_\mathrm{sg}$
being the subgap resistance of the junction. Here, $Q$ is size-independent,
as $R_\mathrm{sg}\propto 1/A$ and $C\propto A$. In the context of MQT,
however, the dynamics takes place at a frequency of $\omega_p$, so that a
complex impedance at that frequency $Z(\omega_p)$ has to be considered. For
an MQT experiment, where the phase and not the charge is the well-defined
quantum variable, the admittance $Y(\omega_p)$ will be responsible for
damping \cite{Ingold_Nazarov}, so that $R$ in (\ref{equ_Qdamp}) will be
given by $R=1/\Re{(Y)}$.

\begin{figure}[t]
\centering
\includegraphics[width=0.8\linewidth,angle=0,clip]{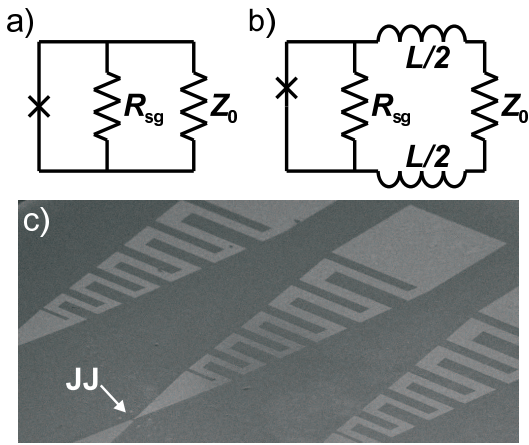}
\caption{\label{impedance}a) Typical impedance environment for switching
experiments in a JJ. As $Z_0\ll R_\mathrm{sg}$, the junction sees the
impedance $Z_0$ at the plasma frequency. b) Lumped element inductors $L$
can be used to decouple the JJ from the line impedance $Z_0$, as discussed
in the text. c) SEM micrograph of the electrode design used for the
investigated junctions.}
\end{figure}

If the junction was an isolated system, the value of $R$ in the context of
MQT would be determined by the intrinsic damping in the zero-voltage state.
The value which is typically taken as a measure for this is the maximal
subgap resistance $R_\mathrm{sg,max}$, which is simply the maximal
resistance value which can be extracted from the nonlinear subgap branch of
the current-voltage characteristics
\cite{Milliken-subgap,Subap_Leakage_Gubrud_2001}. In most experiments
however, the electromagnetic environment of the JJ can be assumed to have
an impedance that is real and accounts for $Z_0\approx 100$~$\Omega$,
corresponding to typical transmission lines \cite{MQT_Martinis_1987}. As
furthermore $Z_0\ll R_\mathrm{sg,max}$ and both contributions are in
parallel (see Fig.\ \ref{impedance}a), we can simply write $Q=\omega_pZ_0
C$ in this case.

Evidently, for junctions having a small capacitance (as in our experiment),
the quality factor $Q=\omega_pZ_0 C$ will be limited to $Q \lesssim 10$ and
additionally depend on the JJ size like $C\propto A$. As we want to
investigate the pure influence of the JJ size on MQT, we would like to
obtain very low damping as well as similar damping for all investigated
junctions. In the implementation of phase qubits, current biased Josephson
junctions have been inductively decoupled from their environment by the use
of circuits containing lumped element inductors and an additional filter
junction \cite{Martinis_first_current_biased_qubit}.  In order to keep our
circuits simple, we attempted to reach a similar decoupling by only using
on-chip lumped element inductors right in front of the JJs (see
Fig.~\ref{impedance}b). This setup leads to an admittance
\begin{equation}\label{YwithL}
Y=1/R_\mathrm{sg,max}+1/(Z_0+i\omega L)\, .
\end{equation}

As for (\ref{YwithL}), we find $\Re{(Y)}\rightarrow 1/R_\mathrm{sg,max}$ in
the limit $\omega L\rightarrow \infty$, big enough lumped element
inductances should decouple the JJ from the $Z_0$ environment and result in
a high intrinsic quality factor $Q=\omega_pR_\mathrm{sg,max}C$ even for
switching experiments. Although it might be difficult to reach this limit
in a real experiment, decoupling inductors should definitively help to
increase the quality factor and move towards a JJ-size independent damping.

The damping in the JJ influences the thermal escape rate (\ref{Gamma_th})
via the prefactor $a_t<1$, which has been calculated for the first time by
Kramers in 1940\cite{Kramers1940}. In the limiting case $Q\rightarrow0$
(moderate to high damping), he found: \begin{equation*}
a_t=\alpha_\mathrm{KMD}=\sqrt{1+\left(\frac{1}{2Q}\right)^2}-\frac{1}{2Q}\,
, \end{equation*} while in the opposite limit $Q\rightarrow\infty$ (very
low damping limit), he found: \begin{equation*}
a_t=\alpha_\mathrm{KLD}=\frac{36\Delta{}U}{5Qk_BT}\, .  \end{equation*}
More recently, B\"{u}ttiker, Harris and Landauer\cite{Buettiker} extended
the very low damping limit to the regime of low to moderate damping finding
the expression\footnote{Equation (\ref{a_t}) does not describe the turnover
from low damping to high damping. This turnover problem has been addressed
by several authors (see e.\,g.\ \citet{Haenggi1990} and references
therein). In general, more precise expressions agree with (\ref{a_t}) in
the parameter regime of our samples to within experimental resolution.}
\begin{equation}\label{a_t}
a_t=\frac{4}{(\sqrt{1+4/\alpha_\mathrm{KLD}}+1)^2}\, .
\end{equation}

Additionally, damping reduces the crossover temperature according to
\cite{Grabert84,Haenggi1990}
\begin{equation}\label{Tcr_Q}
T_\mathrm{cr,Q}=\frac{\hbar\omega_p}{2\pi k_B}\cdot\alpha_\mathrm{KMD}\, .
\end{equation}

A possible way to determine the quality factor $Q$ for such quantum
measurements is to extract it from spectroscopy data
\cite{MQT_Martinis_1987}. Unfortunately, for samples with such a high
critical current density as used in our experiments described here, this
turns out to be experimentally very hard. Hence, we will limit the analysis
of the damping in our experiments to the MQT measurements. However, other
groups have found a good agreement between the $Q$ values determined by
spectroscopy and by MQT \cite{MQT_Martinis_1987,Wallraff2003} and we hope
to observe such a major increase in $Q$ due to the decoupling inductors
that minor uncertainties in $Q$ should not play a role.

\section{Setup and Procedure of Measurement}
\begin{figure}[t]
\centering
\includegraphics[width=\linewidth,angle=0,clip]{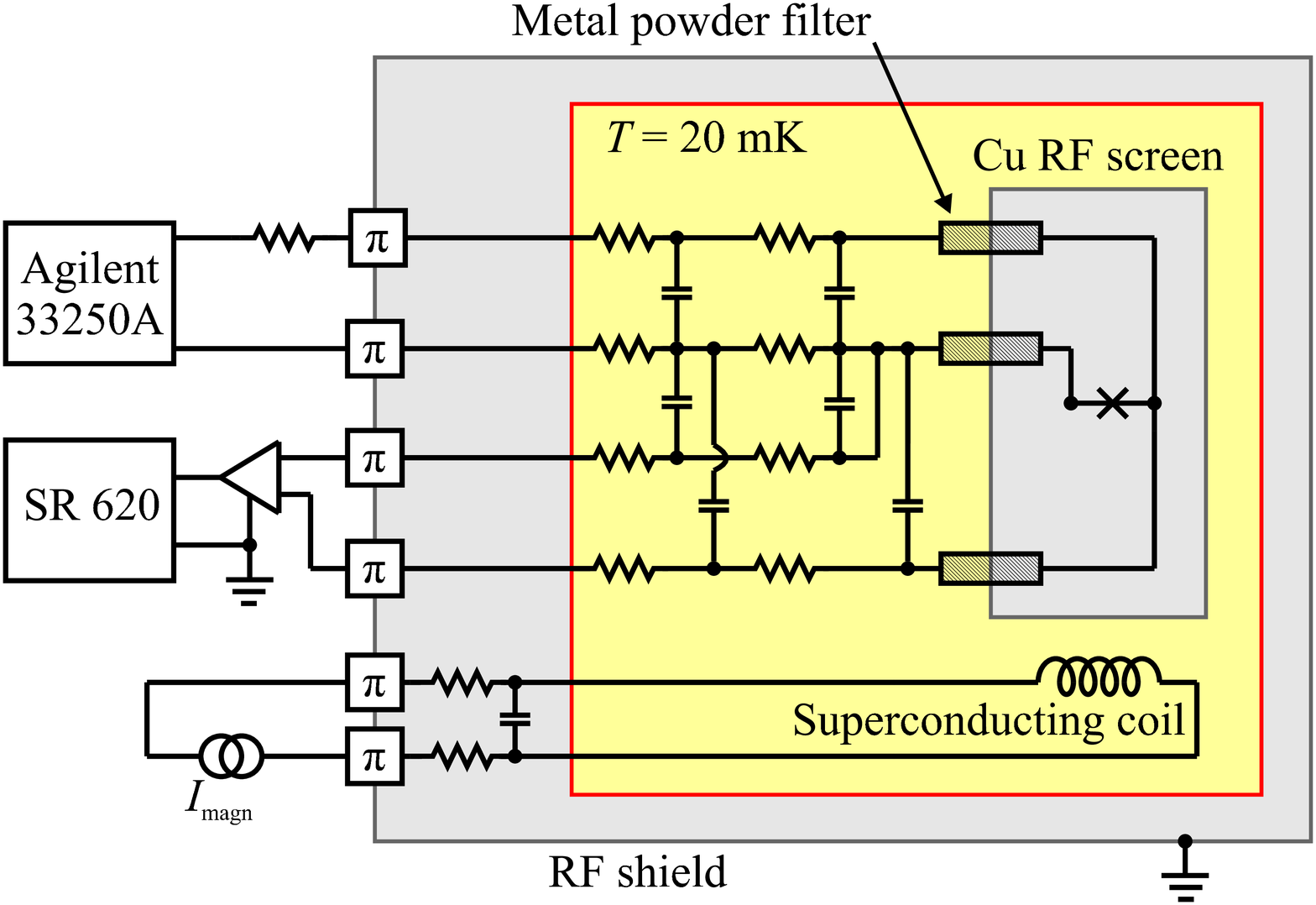}
\caption{\label{figure_messsystem_IFP}Schematic overview of the measurement
system. The superconducting coil and the sample are inside a magnetic
shield consisting of three nested cylindrical beakers, the middle one made
from Pb, the two remaining ones from \textit{Cryoperm}. Furthermore, the
entire dilution refrigerator is placed inside a $\mu$-metal shield at room
temperature. The $\pi$-symbols denote commercial $\pi$-filters.}
\end{figure}
All samples were fabricated by a combined photolithography / electron beam
lithography process based on Nb/Al-AlO$_x$/Nb trilayers. The trilayer
deposition was optimized carefully in order to obtain stress-free Nb films.
For the definition of the Josephson junctions, an Al hard mask is created
employing electron beam lithography. This hard mask acts as an ideal etch
stopper during the JJ patterning with reactive-ion-etching. Furthermore, it
allows the usage of anodic oxidation even for small junctions, which would
not be possible if a resist mask was used. After the anodic oxidation, the
Al hard mask is removed by a wet etching process. Details of this Al hard
mask technique and the entire fabrication process are discussed elsewhere
\cite{Kaiser_fabrication}.

\begin{figure}[t]
\centering \includegraphics[width=\linewidth,angle=0,clip]{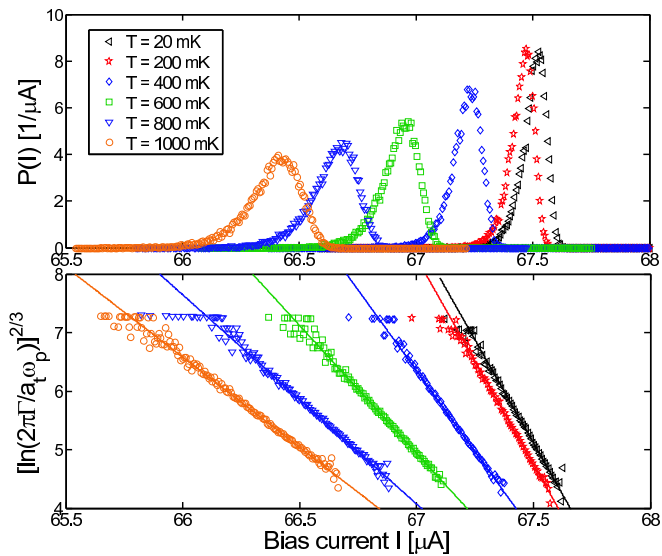}
\caption{\label{PI_Fit}Top: The measured switching current histograms for
sample B3 for selected temperatures. For increasing $T$, the switching
currents decrease and the histograms broaden. Bottom: The plot obtained by
applying (\ref{formulafit}) for the same sample. The fits allow to extract
$T_\mathrm{esc}$ as well as $I_c$.}
\end{figure}

Our measurement setup can be seen in Fig.~\ref{figure_messsystem_IFP}.
Special care has been taken in design of the filtering stages in order to
reach a low-noise measurement environment. The goal of the measurement is
to determine the escape rate $\Gamma$. In order to do so, we have measured
the probability distribution $P(I)$ of switching currents. This was done by
ramping up the bias current with a constant rate $\dot{I}={\rm d}I/{\rm
d}t$ and measuring the time $t_\mathrm{sw}$ between $I=0$ and the switching
to the voltage state with a Stanford Research 620 Counter, so that
$I_\mathrm{sw}=\dot{I}\cdot t_\mathrm{sw}$ could be calculated. An Agilent
33250A waveform generator was used to create a sawtooth voltage signal with
a frequency of 100~Hz, which was converted into the bias current by a
resistor of $47\,{\rm k}\Omega$. In this way, for each temperature,
$I_\mathrm{sw}$ could be measured repeatedly. After doing so 20,000 times,
the switching current histograms $P(I)$ with a certain channel width
$\Delta I$ were attained as shown in the upper part of Fig.~\ref{PI_Fit}.
These histograms were then used to reconstruct the escape rate out of the
potential well as a function of the bias current by employing
\cite{MQT_Martinis_1987,Fulton_D}
\begin{equation}\label{FultonD}
\Gamma(I)=\frac{\dot{I}}{\Delta I} \ln\frac{\sum_{i\geq
I}{P(i)}}{\sum_{i\geq I+\Delta I}{P(i)}}\, .
\end{equation}

With $\Gamma$ at hand, we could now determine the escape temperature
$T_\mathrm{esc}$ by employing (\ref{Gamma_th}). In order to be able to
rearrange this formula, we approximate the potential barrier in the limit
$\gamma\rightarrow 1$ as $\Delta U=4\sqrt{2}/3\cdot
E_J\cdot(1-\gamma)^{3/2}$, so that we find
\begin{equation}\label{formulafit}
\left(\ln\frac{2\pi\Gamma(I)}{a_t(I)\omega_p(I)}\right)^{2/3}=
\left(\frac{4\sqrt{2}E_J}{3k_BT_\mathrm{esc}}\right)^{2/3}
\frac{I_c-I}{I_c}\,.
\end{equation}

Hence, by plotting the left side of (\ref{formulafit}) over the bias
current $I$, we should obtain straight lines (see bottom part of
Fig.~\ref{PI_Fit}). Consequently, we can extract the theoretical critical
current $I_c$ in the absence of any fluctuations as well as the escape
temperature $T_\mathrm{esc}$ by applying a linear fit with slope $a$ and
offset $b$. We then find
\begin{equation*} I_c=-\frac{b}{a}\qquad\mathrm{and}\qquad
T_\mathrm{esc}=-\frac{4\sqrt{2}\Phi_0}{6\pi k_Ba\sqrt{b}}\, .
\end{equation*}
Since $I_c$ enters (\ref{formulafit}) via $E_J$ and $\omega_p$, this
fitting procedure has to be iteratively repeated until the value of $I_c$
converges. So strictly speaking, this procedure involves two fitting
parameters, namely $T_\mathrm{esc}$ and $I_c$. However, it turns out that
$I_c$ is temperature independent within the expected experimental
uncertainty (for all our measurements, the fit values of $I_c$ vary over
the entire temperature range with a standard deviation of only around
0.09~\%). Furthermore, the found $I_c$ values agree very well with the
expected ones from the critical current density $j_c$ of the trilayer and
the junction geometry. Altogether, it can be said that the results for the
main fitting parameter $T_\mathrm{esc}$ should be very reliable.

\section{Sample Characterization}
\begin{figure}[t]
\centering \includegraphics[width=\linewidth,angle=0,clip]{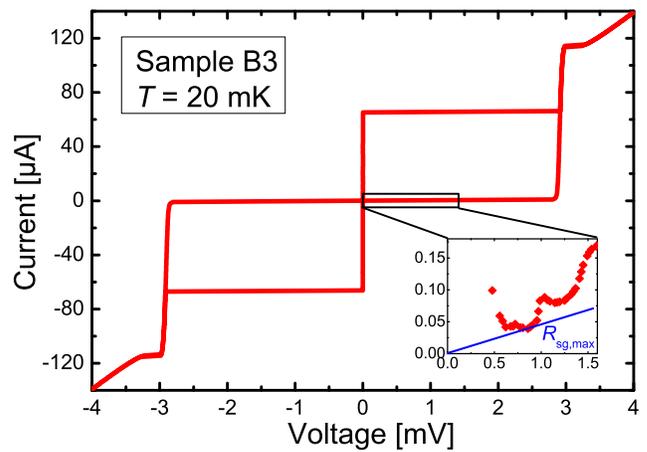}
\caption{\label{char-SampleB3}Characterization of sample B3. The $IV$ curve
shows the high quality regarding the $I_cR_N$ ratio as well as low subgap
currents. The inset shows a magnification of the subgap branch achieved by
a voltage bias. The blue line illustrates how the value for
$R_\mathrm{sg,max}$ was determined. The fact that the current rises for
decreasing voltage at $V\approx 0.5$~V is due to the fact that the junction
jumps back to a supercurrent $I\neq0$ for $V=0$.}
\end{figure}
The JJs were circular in shape and their geometries are given in Table
\ref{table_exp_Tcr}. In order to characterize the samples, $IV$ curves with
current bias as well as $IV$ curves with voltage bias were recorded (an
example can be seen in Fig.~\ref{char-SampleB3}). The quality parameters
for all samples are given in Table~\ref{table_samples_B} and indicate a
very high quality. In the voltage bias measurements, two major current
drops at voltages $2\Delta/2$ and $2\Delta/3$ could be seen and attributed
to Andreev reflections \cite{Arnold-MAR}. Below $2\Delta/3$, we were able
to extract values of the maximal subgap resistance $R_\mathrm{sg,max}$ as
illustrated by the blue line in Fig.~\ref{char-SampleB3}. 

\begin{table}[ht]
\caption{\label{table_samples_B}Experimentally determined parameters for
all investigated JJs. The theoretical critical currents $I_c$ were
extracted from the MQT measurements. The $R_\mathrm{sg,max}$ values were
obtained as shown in the inset of Fig.~\ref{char-SampleB3}a.}
\begin{ruledtabular} \begin{tabular}{ccccc}
Sample & $I_c$ (\textmu{}A) & $V_\mathrm{gap}$ (mV) & $I_cR_N$ (mV) &
$R_\mathrm{sg,max}$ (k$\Omega$)\\\hline
B-1 & $19.1$ & 2.88 & $1.75$ & $54.0$ \\
B-2 & $31.9$ & 2.88 & $1.86$ & $73.0$ \\
B-3 & $68.1$ & 2.92 & $1.93$ & $21.1$ \\
B-4 & $70.8$ & 2.90 & $1.91$ & $31.5$ \\
\end{tabular} \end{ruledtabular}
\end{table}

\section{Results and Discussion}
\subsection{Damping in the Junctions}
In order to decouple the JJs from their environmental impedance, the
electrodes leading to the junctions were realized as lumped element
inductors, as can be seen in Fig.~\ref{impedance}c. This design was based
on the layout that we recently used to successfully realize lumped element
inductors for $LC$ circuits in the GHz frequency range
\cite{Kaiser_Dielectric_Losses}. Furthermore, simulations with Sonnet
\footnote{Sonnet Software Inc., 1020, Seventh North Street, Suite 210,
Liverpool, NY 13088, USA} confirmed that the meandered electrodes indeed
act as lumped element inductors at the relevant frequencies
$\omega_p(\gamma_\mathrm{cr})$. The complex simulation with Sonnet gives an
inductance of $L/2\approx 1.65\,$nH (for one electrode) while the much
simpler analysis with FastHenry\footnote{Fast Field Solvers,
http://www.fastfieldsolvers.com} yields $L/2\approx 1.8\,$nH.

\begin{figure}[t]
\centering \includegraphics[width=\linewidth,angle=0,clip]{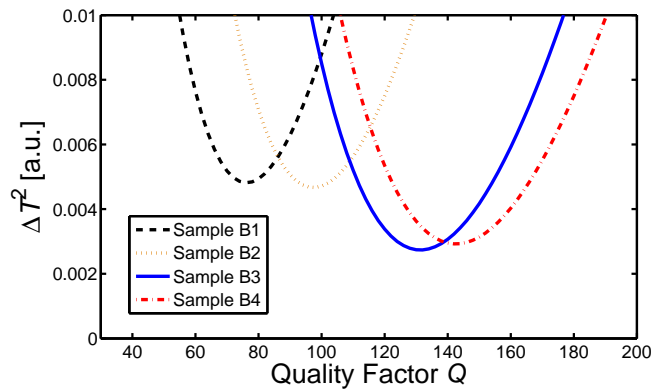}
\caption{\label{Fig_LeastSquares}Determination of the experimentally
observed quality factor $Q$ for all samples. The curves are minimal when
the determined $T_\mathrm{esc}$ values deviate the least from the
corresponding bath temperatures $T$ in the thermal regime $T>500$~mK.}
\end{figure}

For each sample, the data were analyzed using a number of different $Q$
values in order to see if we could determine the experimentally observed
damping. This was done by calculating the deviation of $T_\mathrm{esc}$
from the bath temperature $T$ in the thermal regime:
\begin{equation}\label{form_leastsquares}
\Delta T^2 = \sum_{T>500\,{\rm mK}}\left(T_\mathrm{esc}-T\right)^2
\end{equation}
and finding its minimum value regarding $Q$. The corresponding values were
then used for the sample analysis. It can be seen in
Fig.~\ref{Fig_LeastSquares} that the points of experimentally observed
damping could be clearly identified. The evaluated $Q$ values are given in
Table~\ref{table_MQT_damping}.

In a preliminary experiment, we investigated MQT in a junction with a
diameter of $d=1.9$~\textmu{}m, a critical current of $I_c\approx
12$~\textmu{}A and low-inductance electrodes, which were simple wide lines
and can be imagined as the envelope of the electrodes in
Fig.~\ref{impedance}c.  We carried out a similar analysis to determine the
damping and obtained a quality factor of $Q=4$. Subsequently, we evaluated
(\ref{equ_Qdamp}) and calculated an impedance of $R = 99.8$~$\Omega$, which
is very close to the expected value of $Z_0\approx100$~$\Omega$ for typical
transmission lines \cite{MQT_Martinis_1987}. This means that with this
simple preliminary design, the junction was in no way decoupled from the
electromagnetic environment.

The $Q$ values obtained by using inductive electrodes (see
Tab.~\ref{table_MQT_damping}), however, show that we have drastically
increased the quality factors with respect to the preliminary measurement.
If we calculate the $R$ values using (\ref{equ_Qdamp}), we find that they
are clearly above the typical line impedance of $Z_0\approx100$~$\Omega$ as
well as the vacuum impedance of 377~$\Omega$, which shows that we were
indeed able to inductively decouple the JJ from its usual impedance
environment. As expected, the determined $R$ values are still clearly below
the subgap resistance $R_\mathrm{sg,max}$, indicating that we have not
reached the limit $\omega L\rightarrow \infty$.  Instead, we are in the
intermediate regime $Z_0\ll R\ll R_\mathrm{sg,max}$, leading to the fact
that $Q$ still exhibits a slight dependence on the JJ size (see
Table~\ref{table_MQT_damping}). However, all JJs are in the low-damping
regime, so that no influence of damping on the results should be present
and differences in the experimental results should indeed be due to the JJ
size. This can be seen by the fact that the damping related correction in
$T_\mathrm{cr}$ according to equation (\ref{Tcr_Q}) is smaller than 1~\%
for all experimentally observed $Q$ values. Altogether, we can state that
we will be able to carry out our investigation of the size dependence of
MQT with very low and nearly size-independent damping.

\begin{table}[b]
\caption{\label{table_MQT_damping}The experimentally determined values
characterizing the damping for all samples. The $L_\mathrm{calc}$ values
were determined from $Q$, the values given in Table \ref{table_samples_B}
and equation (\ref{YwithL}). They are in good agreement with the design
value of $L\approx 3.3\,$nH.}
\begin{ruledtabular} \begin{tabular}{cccc}
Sample & Q & $R$ ($\Omega$) & $L_\mathrm{calc}$ (nH) \\ \hline
B-1 & $76$ & $1571$ & $1.33$ \\
B-2 & $98$ & $1327$ & $1.39$ \\
B-3 & $132$ & $1015$ & $1.36$ \\
B-4 & $143$ & $1041$ & $1.43$ \\
\end{tabular} \end{ruledtabular}
\end{table}

In addition to the rather qualitative considerations above, we performed a
quantitative analysis employing equation (\ref{YwithL}). If we use
$\omega=\omega_p(\gamma_\mathrm{cr})$, take $R_\mathrm{sg,max}$ from
Table~\ref{table_samples_B} and assume that $Z_0=100\,\Omega$, we can
calculate the decoupling inductance $L_\mathrm{calc}$ for all samples. The
values, given in Table~\ref{table_MQT_damping}, are a factor of around
$2.3-2.5$ smaller than the simulation value of $L\approx 3.3\,$nH, but of
the right order of magnitude. For such a complex system, this is a
surprisingly good agreement between simulation and theory on the one side
and experimentally determined values on the other side. In summary, we
conclude that we have successfully demonstrated that decoupling of the
Josephson junction from its environment is also possible using only lumped
element inductors.

\subsection{Crossover to the Quantum Regime}
\begin{figure}[t]
\centering
\includegraphics[width=\linewidth,angle=0,clip]{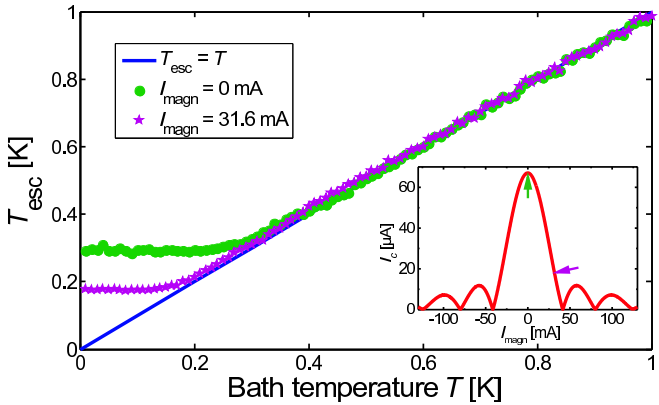}
\caption{\label{Fig_MQT_magnet}Calculated escape temperatures for sample B3
with and without an applied magnetic field. The inset shows the $I_c(\Phi)$
modulation of this junction; the arrows indicate where the MQT data was
obtained. For $I_\mathrm{magn}=31.6\,$mA, a clear reduction of the observed
crossover temperature $T_\mathrm{cr}$ is observed.}
\end{figure}
We now turn to the investigation of the crossover point from the thermal to
the quantum regime and the influence of JJ size on it. As can be seen in
Table~\ref{table_exp_Tcr}, we expect a clear reduction of $T_\mathrm{cr}$
with increasing JJ size. However, an experimental observation of lower
crossover temperatures for smaller JJs having smaller critical currents
could simply be due to current noise in our measurement setup. In order to
exclude this, we artificially reduced the critical current of sample B3 by
applying a magnetic field in parallel to the junction area. While unwanted
noise should now lead to an increase in the observed $T_\mathrm{cr}$, the
physical expectation is a significantly reduced $T_\mathrm{cr}$ due to the
lower plasma frequency according to (\ref{Tcr}). The result of this
measurement can be seen in Fig.\ \ref{Fig_MQT_magnet} and
Table~\ref{table_MQT_results}. We found an agreement between calculated and
observed crossover temperature down to $T_\mathrm{cr}\approx 140$~mK, which
was the lowest temperature we examined. Hence, it is clear that we have a
measurement setup exhibiting low noise, where the electronic temperature is
indeed equal to the bath temperature. The lowest investigated temperature
of 140~mK is clearly below any temperature needed for the comparison of the
JJs of different sizes with each other.

\begin{figure}[t]
\centering \includegraphics[width=\linewidth,angle=0,clip]{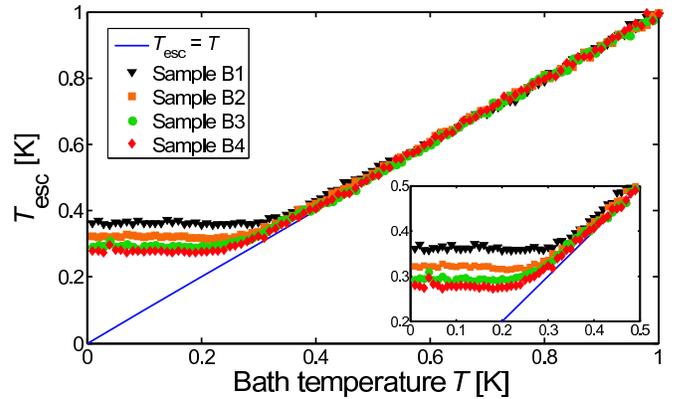}
\caption{\label{Fig_MQT_sizes}Calculated escape temperatures for all
samples.  The crossover to the quantum regime is very clear in each
measurement. The inset shows a magnification of the quantum regime.  The
reduction of the crossover temperature $T_\mathrm{cr}$ with increasing JJ
size is clearly visible.}
\end{figure}

Finally, we measured the switching histograms for the four JJs of different
sizes and evaluated the escape temperature $T_\mathrm{esc}$ and the
theoretical critical current $I_c$. This allowed us to determine the
crossover temperature $T_\mathrm{cr}$ and the normalized crossover current
$\gamma_\mathrm{cr}=I_\mathrm{sw,cr}/I_c$. We indeed found a clear
dependence of the crossover temperature on the JJ size as can be seen in
Fig.~\ref{Fig_MQT_sizes}. To compare the experimental $\gamma_\mathrm{cr}$
and $T_\mathrm{cr}$ values with the ones expected by theory, we now
performed the theoretical calculation described above using the
experimentally determined $Q$ values and equation (\ref{Tcr_Q}). All
experimentally determined values are in excellent agreement with theory, as
can be seen in Table~\ref{table_MQT_results}. 

\begin{table}[ht]
\caption{\label{table_MQT_results}The experimentally determined values
characterizing the crossover from the thermal to the quantum regime in
comparison with the theoretical expectations for all measurements.}
\begin{ruledtabular} \begin{tabular}{cdcccc}
Sample & \multicolumn{1}{c}{$I_c(\Phi)/I_c(0)$} & $\gamma_\mathrm{cr,theo}$
& $\gamma_\mathrm{cr,exp}$ & $T_{\mathrm{cr,}Q\mathrm{,theo}}$ &
$T_\mathrm{cr,exp}$ \\
& & & & (mK) & (mK) \\ \hline
B-1 & 1 & $0.965$ & $0.970$ & $368$ & $362$ \\
B-2 & 1 & $0.977$ & $0.981$ & $322$ & $321$ \\
B-3 & 1 & $0.988$ & $0.990$ & $290$ & $294$ \\
B-3 & 0.52 & $0.984$ & $0.987$ & $223$ & $236$ \\
B-3 & 0.27 & $0.979$ & $0.983$ & $172$ & $176$ \\
B-3 & 0.13 & $0.973$ & $0.975$ & $129$ & $147$ \\
B-4 & 1 & $0.988$ & $0.990$ & $276$ & $278$ \\
\end{tabular} \end{ruledtabular}
\end{table}

\section{Conclusions}
We have carried out systematic Macroscopic Quantum Tunneling (MQT)
experiments with varying Josephson junction area. Our samples were
fabricated on the same chip. Thorough characterization before the actual
quantum measurements revealed that the junctions exhibit a very high
quality. We showed that we could significantly decrease the damping at
frequencies relevant for MQT by using lumped element inductors, which
allowed us to perform our study in the low damping limit. The crossover
from the thermal to the quantum regime was found to have a clear and
systematic dependence on junction size, which is in perfect agreement with
theory.

\begin{acknowledgments}
This work was partly supported by the DFG Center for Functional
Nanostructures, project number B1.5. We would like to thank A. V. Ustinov
for useful discussions. 
\end{acknowledgments}

\bibliography{MQT_PRB}

\end{document}